\documentclass[11pt,
showpacs,
preprintnumbers,amsmath,amssymb,
aps,prd,nofootinbib,eqsecnum,a4paper]{revtex4-1}
\usepackage{latexsym}
\usepackage{amsmath}
\usepackage[utf8x]{inputenc}
\usepackage[T1]{fontenc}
\usepackage{subdepth}
\usepackage{color}
\usepackage{soul}
\usepackage{ulem}
\begin{document}

\title{Noncommutative SUSY Black Holes}

 \author{A. Crespo-Hern\'andez}
\email{andres.crespo@academicos.udg.mx}
\affiliation{ Preparatoria 5, Universidad de Guadalajara
Av. Fray Andr\'es de Urdaneta s/n, C.P. 44930, Guadalajara, Jalisco, M\'exico. }%

\author{E. A. Mena-Barboza}
\email{eri.mena@academicos.udg.mx}
\affiliation{ Centro Universitario de la Ci\'enega, Universidad de Guadalajara,Ave. Universidad 1115 Ed. de Tutor\'{\i}as e Investigaci\'on, C.P. 47820 Ocotl\'an, Jalisco, M\'exico. }%

\author{M. Sabido}
\email{msabido@fisica.ugto.mx}
\affiliation{ Departamento  de F\'{\i}sica de la Universidad de Guanajuato,\\
 A.P. E-143, C.P. 37150, Le\'on, Guanajuato, M\'exico
 }%

\begin{abstract}
In this paper we propose a generalization to the Schwarzschild metric and define noncommutative SUSY  black holes. We introduce the noncommutative deformation to the minisuperspace variables and derive the noncommutative  supersymmetric (SUSY) Wheeler-DeWitt (WDW) equation for the  Schwarzschild black hole. {We calculate the metric and find that the singularities are not removed.}
 \end{abstract}

\maketitle
\section{Introduction}
 
{Black Holes  are one of the most enigmatic and fascinating objects in physics. It has been a very active  area of research, and due to }
the spectacular observations from the LIGO \cite{ligo}  and  the Event Horizon collaborations \cite{EH}, there has been 
a resurgence in the study of theoretical puzzles in black hole physics.
Moreover, because there are black hole solutions in alternative theories of gravity, black holes are used as workhorses to probe fundamental aspects of these theories (i.e supergravity, noncommutative gravity, Hordenski gravity, etc.).

{The idea of a noncommutative space-time was revived at the beginning of the century. There have been several attempts to study the possible effects of noncommutativity in the cosmological scenario. In previous works\cite{tachy,yee,yee2}, it has been argued that there is a possible relationship between the  cosmological constant and the noncommutative parameters. Moreover, the effects of noncommutativity during inflation were explored\cite{brand}, but noncommutativity was only incorporated in the matter Lagrangian, neglecting the gravitational sector. The difficulties of analyzing noncommutative gravitational models arise from the complicated structure of noncommutative gravity. Writing a
noncommutative theory of gravity which only depends on the
commutative fields and their derivatives has  field equations that are
highly nonlinear \cite{sabidograv1,sabidograv2}. To avoid these difficulties, it has been
proposed to introduce the effects of noncommutativity at the quantum level\cite{ncqc},
namely quantum cosmology, by deforming the minisuperspace through a Moyal deformation of the Wheeler-DeWitt (WDW) equation. It is then possible to proceed in noncommutative quantum mechanics \cite{gamboa}.} Using this approach for noncommutative cosmology and using the diffeomorphism to transform the
Schwarzschild metric into the Kantowski-Sachs (KS) metric, one obtains the noncommutative Wheeler-DeWitt (NC-WDW) equation
for the Schwarzschild black hole\cite{LopezDominguez:2007zz}. From the NC-WDW equation and the Feynman-Hibbs method,  they calculate the entropy of the noncommutative black hole. This method was originally used for the Schwarzschild black hole, to reproduce the known results \cite{paths}.

We know that supergravity is the  supersymmetric generalization of general relativity (GR).
Supersymmetric black hole solutions of supergravity theories played a crucial role in important developments in string black hole physics.
{In supergravity, only black holes that satisfy the BPS constraints are well understood.\footnote{{The simplest case for a spherical symmetric solution that meets this criteria is the  Reissner-Nordstr$\ddot{o}$m extremal black hole.}} Consequently non BPS black holes have not been extensively studied.}

{One approach that has been used to study SUSY black holes is to use the relationship between the KS and the Schwarzschild metrics, to introduce supersymmetry}. Along this line of reasoning,  classical (and quantum) supersymmetric Schwarzschild and Schwarzschild-(anti) de Sitter black hole models were proposed \cite{julio1,julio2}. By  supersymmetrizing the WDW  equation associated with the standard Schwarzschild black hole. {The authors derive a modified (SUSY quantum) Hamiltonian and its corresponding classical equations that in this sense define a supersymmetric generalization of the  Schwarzschild and Schwarzschild-(anti) de Sitter space-times.} 

 {Exploiting the relationship between the KS and Schwarzschild metric
 we can introduce new physical ideas to black holes. It was effective to introduce noncommutative effects using the WDW equation allowing us to construct noncommutative Schwarzschild black hole. It is reasonable to assume that noncommutative constructions which deal with the supersymmetric versions of the WDW equation have a similar behaviour \cite{eri_andres}, this motivate us to study the effects of noncommutativity on SUSY black holes, using the NC-SUSY WDW equation, therefore the main objective of this paper is to explore black holes in the context of  noncommutativity and supersymmetry.}
 
The paper is organized as follows. In section \ref{sec_1},  we review the proposal for SUSY black holes \cite{julio1}, we also discuss noncommutative black holes. In Section \ref{sec_3} we present our proposal for noncommutative SUSY black Holes. Section \ref{final} is devoted to discussion and final remarks.

\section{Modifying the Wheeler-DeWitt equation}\label{sec_1}

{In this section we discuss how to introduce the new physics to the Schwarzschild  black hole. We will briefly discuss SUSY black holes from the WDW equation. We also discuss the introduction of noncommutativity to black holes by deforming the WDW equation.}

\subsection{{Wheeler-DeWitt equation and the SUSY black hole}}

{Let us start by recalling} the classical and quantum aspects of the  supersymmetric cosmological KS model and  the Schwarzschild black hole. We use the square root  and operator method. Where the resulting Hamiltonian has terms that allow us to find a supersymmetric solution \cite{julio1}. From the Schwarzschild metric 
\begin{equation}
ds^{2}=-\left(1-\frac{2M}{r}\right)dt^{2}+\left( 1-\frac{2 M}%
{r}\right)^{-1}dr^{2}
+r^{2}\left(d \theta^{2}+\sin^{2}\theta d\varphi^{2}\right),
\end{equation}
we can find the relationship between the cosmological Kantowski-Sachs metric and the Schwarzschild metric, 
by doing the coordinate transformation $t\leftrightarrow r$. Also, $g_{tt}$ and $g_{rr}$ change their sign and $\partial_{t}$ becomes a space-like vector. {Finally, we compare the Kantowski-Sachs metric with the parametrization by Misner and identify}%

\begin{equation}
N^{2}=\left(\frac{2M}{t}-1\right)^{-1},\quad
e^{2\sqrt{3}\beta}=\frac{2M}{t}-1,\quad
e^{ -2\sqrt{3}\beta}e^{-2\sqrt{3}\Omega}=t^{2},
\label{dif}
\end{equation}
where we know that 
\begin{equation}
ds^{2}=-N^{2}dt^{2}+e^{2\sqrt{3}\beta} dr^{2}+
e^{-2\sqrt{3}\beta}
 e^{\left(  -2\sqrt{3}\Omega\right)}  \left(
d\theta^{2}+\sin^{2}\theta d\varphi^{2}\right).
\label{KSmetric}%
\end{equation}
{We canonically quantize this model and get the} Wheeler-DeWitt equation for the
Kantowski-Sachs metric. {Which, with some particular factor ordering, is given by}
\begin{equation}
\left[  -\frac{\partial^{2}}{\partial\Omega^{2}}+\frac{\partial^{2}}%
{\partial\beta^{2}}+48e^{ -2\sqrt{3}\Omega } \right]
\psi(\Omega,\beta)=0.\label{ks}
\end{equation}
{To construct the supersymmetric generalization for the WDW equation Eq.(\ref{ks}), we follow} the procedure for SUSY quantum cosmology\cite{julio1}. {First we need to find a diagonal Hamiltonian operator constructing the supercharges and get the supersymmetric KS WDW equation}
\begin{equation}
\left[-\frac{\partial^{2}}{\partial\Omega^{2}}+\frac{\partial^{2}}%
{\partial\beta^{2}}+12e^{-2\sqrt{3}\Omega}(4\pm e^{\sqrt{3}\Omega})\right]
\psi_{\pm}(\Omega,\beta)=0.\label{kssusy}
\end{equation}
{For the metric, we apply the WKB method and from the semiclassical equivalent to  Eq.(\ref{kssusy})}, we find
\begin{align}
4e^{-2\sqrt{3}\Omega}\left(1+2t\sqrt{3}\dot{\Omega}\right)-t^2
\left(4\pm e^{\sqrt{3}\Omega}\right)=0,\label{clasusy}
\end{align}
for our purposes, the only physically relevant asymptotic region of Eq.(\ref{clasusy})  is $4\ll e^{\sqrt{3}\Omega}$.
From the solutions for the asymptotic regions 
we get
\begin{align}
e^{-\sqrt{3}\Omega}=\left(\frac{3}{4}\right)^{1/3}r^{2/3}
\left(\pm1+\frac{C}{\sqrt{r}}\right)^{1/3},\label{solposneg}
\end{align}
where $C$ is a constant. Using these solution and Eq.(\ref{KSmetric}) we construct the metric
\begin{align}
ds^{2}=&-\left(\frac{3}{4}\right)^{2/3}r^{-2/3}
\left(\pm1+\frac{C}{\sqrt{r}}\right)^{2/3}dt^{2}+
\left(\frac{4}{3}\right)^{2/3}r^{2/3}\left(\pm1+\frac{C}{\sqrt{r}}\right)^{-2/3}
dr^{2}\nonumber\\
&+r^{2} \left(d \theta^{2}+\sin^{2} \theta
d\phi^{2}\right).\label{solucionposneg}
\end{align}
which is formally equivalent to Eq.(\ref{KSmetric}). {Note that there are two cases in Eq.(\ref{solucionposneg}), it was shown that the interesting and physical case is for $C>0$}. For this case, there are two singularities in $r=0$ and $r=C^2$.
{Moreover, it was suggested that because the semiclassical limit of Eq.(\ref{kssusy}) contains the ``fermionic" information, consequently, it is reasonable to expect the lack of a horizon. This was analyzed by solving the Dirac equation in the Schwarzschild and Kerr backgrounds and determining that the spinors destroy the  horizon\cite{gibbons}.  }

\subsection{Noncommutative Black Hole}\label{sec_2}

The starting point for the analysis is to  consider the noncommutative proposal of quantum cosmology \cite{ncqc}. Where the Cartesian coordinates $\Omega$ and $\beta$ of the KS minisuperspace variables are introduced as a canonical deformation in the algebra of the minisuperspace operators,

\begin{equation}
[\hat{\Omega},\hat{\beta}]=\vartheta,\quad
[\hat{\Omega},\hat{P_\Omega}]=[\hat{\beta},\hat{P_\beta}]=1,\quad [\hat{P_\Omega},\hat{P_\beta}]=0. \label{ncc}
\end{equation}
Given the Weyl quantization procedure in the context of the above description, the realization of the commutation relation Eq.(\ref{ncc}) between the minisuperspace variables is made by a specific Moyal product 

\begin{equation}
f(\Omega,\beta)\ast g(\Omega,\beta)=f(\Omega,\beta) e^{(\frac{i\vartheta}{2})(\stackrel{\leftarrow}{\partial}_{{\Omega}}\stackrel{\rightarrow}{\partial}_{\beta} -\stackrel{\leftarrow}{\partial}_{\beta}\stackrel{\rightarrow}{\partial}_{\Omega})} g(\Omega,\beta).
\end{equation}
This leads to a shift in the variables
\begin{equation}
\hat{\Omega}=\Omega-\frac{\vartheta}{2}P_\beta,\quad \hat{\beta}=\beta+\frac{\vartheta}{2}P_\Omega. \label{shift}
\end{equation}
The Moyal product functions will be applied to find the WDW equation, this gives a modified
WDW equation for the noncommutative model%
\begin{equation}
\left[-P_{\Omega}^{2}+P_{\beta}^{2}-48e^{-2\sqrt{3}\Omega}\right]
\ast\psi(\Omega,\beta)=0.
\end{equation}
As is known in noncommutative quantum mechanics, the original
phase-space is modified. It is possible to reformulate in terms
of the commutative variables and the ordinary product of functions, if
the new variables satisfy Eq.(\ref{shift}).  Consequently, the
original WDW equation changes, with a modified
potential  $V(\Omega,\beta)$,
\begin{equation}
V\left( \Omega,\beta\right) \ast\psi\left( \Omega,\beta\right) =V\left(
\Omega-\frac{\vartheta}{2} P_{\beta},\beta+\frac{\vartheta}{2}P_{\Omega}\right)\psi\left(\Omega,\beta\right),
\end{equation}
so the NC-WDW equation takes the form%
\begin{equation}
\left[  -\frac{\partial^{2}}{\partial\Omega^{2}}+\frac{\partial^{2}}%
{\partial\beta^{2}}+48e^{\left( -2\sqrt{3}\Omega+\sqrt{3}\vartheta
P_{\beta}\right) } \right]  \psi(\Omega,\beta)=0. \label{ncwdw}
\end{equation}
{The consequences of the NC-WDW equation were originally analyzed in cosmology \cite{ncqc}. At the quantum level, it gives several new maxima on the probability density depending on the value of the noncommutative parameter $\vartheta$. The classical solutions have been obtained for the model \cite{bastos1}. Moreover, using the NC-WDW in Eq.(\ref{ncwdw}) with the Feynman-Gibbs approach to statistical mechanics,  the thermodynamics of noncommutative black holes were studied \cite{LopezDominguez:2007zz}. Moreover, using Eq.(\ref{ncwdw}), the singularity of noncommutative quantum black holes was discussed\cite{bastos2}.}

\section{Noncommutative SUSY Black Hole}\label{sec_3}

As already stated in the previous sections,  to  construct the noncommutative SUSY black hole we will start with the SUSY WDW equation for the Kantowski-Sachs cosmological model Eq.(\ref{kssusy}). 
This gives the noncommutative SUSY WDW equation. After introducing the deformed algebra in Eq.(\ref{ncc}) we get a SUSY generalization of the Schwarzschild black hole
\begin{equation}
\left[  -\frac{\partial^{2}}{\partial\Omega^{2}}+\frac{\partial^{2}}%
{\partial\beta^{2}}+12e^{-2\sqrt{3}(\Omega-\frac{\vartheta}{2}P_\beta)}(4\pm e^{\sqrt{3}(\Omega-\frac{\vartheta}{2}P_\beta)})\right] \psi(\Omega,\beta)=0. \label{susyncwdw}
\end{equation} 
Then we apply the WKB method to the NC SUSY WDW equation. Assuming that the wave function has the form
\begin{equation}
	\psi(\Omega,\beta)=e^{i(S_1(\Omega)+S_2(\beta))},
\end{equation}
we can construct the Einstein-Hamilton-Jacobi (EHJ) equation. Finally one can derive the equation of motion\cite{julio1}. This approach is equivalent to using a modified Hamiltonian that includes the noncommutative deformation as well as the SUSY generalization. 

From the EHJ equation one can identify $\frac{dS_1(\Omega)}{d\Omega}\rightarrow P_\Omega$ and $\frac{dS_1(\beta)}{d\beta}\rightarrow P_\beta$, 
where $P_\Omega$ and $P_\beta$ are the conjugate momentum to $\Omega$ and $\beta$, respectively. 

We can also  consider the noncommutative relations Eq.(\ref{ncc}) and the shift in the noncommutative variable Eq.(\ref{shift}) in Eq.(\ref{kssusy}). From the noncommutative SUSY WDW equation we can write the Hamiltonian as
\begin{equation}
	H=P_\Omega^2-P_\beta^2+12e^{-2\sqrt{3}(\Omega-\frac{\vartheta}{2}P_\beta)}(4\pm e^{\sqrt{3}(\Omega-\frac{\vartheta}{2}P_\beta)}).\label{hnc2}
\end{equation}
This is the same Hamiltonian we obtain from Eq. (\ref{susyncwdw}).
{ We are interested in asymptotic region where supersymmetry can dominate \cite{julio1}. Therefore, we take the approximation}  $e^{\sqrt{3}\Omega}\gg 4 $. {This can be considered the ``classical supersymmetric" limit}. In this limit, the equations of motion are
\begin{eqnarray}
\dot{\Omega}&=&-\frac{e^{2\sqrt{3}(\Omega-\frac{\vartheta}{2}P_\beta)}P_\Omega}{12},\quad
\dot{P}_\Omega=\mp\frac{\sqrt{3}}{2}e^{\sqrt{3}(\Omega-\frac{\vartheta}{2}P_\beta)}\\
\dot{\beta}&=&\mp\frac{\sqrt{3}}{4}\vartheta e^{\sqrt{3}(\Omega-\frac{\vartheta}{2}P_\beta)}+\frac{e^{2\sqrt{3}(\Omega-\frac{\vartheta}{2}P_\beta)}}{12}P_\beta,\quad
\dot{P_\beta}=0.\nonumber
\end{eqnarray}
{As one expects that the effects of noncommutativity are very small, and to simplify the calculations, we take the approximation $\vartheta\ll 1$. Moreover, one can obtain definitions for the momentum from the Hamiltonian}
\begin{eqnarray}
P_\Omega&=&-12e^{-2\sqrt{3}\Omega}\dot{\Omega}-144\sqrt{3}\vartheta e^{-4\sqrt{3}\Omega}\dot{\beta}\dot{\Omega},\label{poc2a}\\
P_\beta&=&\pm3\sqrt{3}\vartheta e^{-\sqrt{3}\Omega}+12e^{-2\sqrt{3}\Omega}\dot{\beta}+144\sqrt{3}\vartheta e^{-4\sqrt{3}\Omega}\dot{\beta}^2.\label{pbc2a}
\end{eqnarray}
From the mentioned approximations we get
\begin{eqnarray}
&-&3\dot{\Omega}^2-72\sqrt{3}\vartheta e^{-2\sqrt{3}\Omega}\dot{\beta}\dot{\Omega}^2\mp\frac{3\sqrt{3}}{2}\vartheta e^{\sqrt{3}\Omega}\dot{\beta}+3\dot{\beta}^2\nonumber\\
&+&72\sqrt{3}\vartheta e^{-2\sqrt{3}\Omega}\dot{\beta}^3\mp\frac{1}{4} e^{3\sqrt{3}\Omega}=0.\label{edbo2}
\end{eqnarray}
Setting  $t^2=e^{-2\sqrt{3}\Omega-2\sqrt{3}\beta}$ and using the change of variable $u=e^{-3\sqrt{3}\Omega}$, we get
\begin{eqnarray}
\mp\frac{9}{4}\vartheta t^{-3/2} u^{2/3}\pm\frac{3}{4}\vartheta t^{-1/2} u^{-1/3}\dot{u}\pm\frac{3}{8}t^{-1/2}-\frac{3}{2}t^{-5/2}u+t^{-3/2}\dot{u}\nonumber\\
+36\vartheta t^{-7/2}u^{5/3}-36\vartheta t^{-5/2} u^{2/3}\dot{u}+8\vartheta t^{-3/2}u^{-1/3}\dot{u}^2=0.
\end{eqnarray}	
To solve this equation, we use a perturbative method. We start by proposing that $u=\pm\frac{3}{4}t^2+ct^{3/2}+\vartheta u_1+O(\vartheta^2)$. The first and second terms are the same as for the SUSY black hole\cite{julio1}. Substituting, we  obtain

\begin{equation}
t^{-3/2} \dot u_1-\frac{3}{2}t^{-5/2}u_1=\pm\frac{27 \left(2 c\pm \sqrt{t}\right)}{8 \sqrt[3]{8 c t^{3/2}\pm 6 t^2}}.
\end{equation}
After solving for $t$  and interchanging $r\leftrightarrow t$, we arrive  at
\begin{align}
e^{-\sqrt{3}\Omega}=\left(\frac{3}{4}\right)^{1/3}r^{2/3}\left[\pm1+Cr^{-1/2}+\frac{9}{40}\vartheta r^{-1/2} \left(9C\pm 4\sqrt{r}\right) \left(6C \pm6  \sqrt{r}\right)^{2/3}\right]^{1/3}\\
e^{-\sqrt{3}\beta}=\left(\frac{4}{3}\right)^{1/3}r^{1/3}\left[\pm1+Cr^{-1/2}+\frac{9}{40}\vartheta r^{-1/2} \left(9C\pm 4\sqrt{r}\right) \left(6C \pm6  \sqrt{r}\right)^{2/3}\right]^{-1/3}\label{6}
\end{align}
Finally, substituting the solutions in Eq.(\ref{KSmetric}), we find the noncommutative SUSY generalization to the Schwarzschild  metric
\begin{eqnarray}
ds^2&=&-\left(\frac{3}{4}\right)^{2/3}r^{-2/3}\left[\pm1+\frac{C}{\sqrt{r}}+\frac{9\vartheta\left(9C\pm 4\sqrt{r}\right) \left(6C \pm6  \sqrt{r}\right)^{2/3}}{40\sqrt{r}} \right]^{2/3}dt^2\nonumber\\
&+&\left(\frac{4}{3}\right)^{2/3}r^{2/3}\left[\pm1+\frac{C}{\sqrt{r}}+\frac{9\vartheta\left(9C\pm 4\sqrt{r}\right) \left(6C \pm6  \sqrt{r}\right)^{2/3}}{40\sqrt{r}} \right]^{-2/3}dr^2\nonumber\\
&+&r^2(d\theta^2+\sin{\theta}d\phi^2).
\end{eqnarray}
{These solutions are valid only in the asymptotic region $e^{\sqrt{3}\Omega}>>4$.  The different solutions to reconstruct the metric give rise to some branches that are not physical as well as some singularities, these are summarized in table I.

\noindent We can see that for $\vartheta=0$ one recovers the results for the SUSY black Hole \cite{julio1}.} 
\begin{table}[h!]
\caption{Singular and regular points for the noncommutative SUSY metric.}
{\begin{tabular}{|c | c | c | c|} 
\hline
\textbf{$\pm$ 1} &  $\mathbf{r}$ &  $\mathbf{C}$&\textbf{Type of region}\\
\hline
\hline
Positive/Negative & $r=0$  & Any value & Singular point \\
\hline
\hline
Positive & $r>0$ &  $C>0$ & {Regular region}\\
      \hline
      Positive & $r=C^2$ &  $C<0$ & {Singular point}\\
      \hline
      \hline
      Negative & $r>0$ &  $C<0$ & {Invalid solution for the metric}\\
      \hline
      Negative & $r=C^2$ &  $C>0$ & {Singular point}\\
      \hline
\end{tabular} \label{table1}}
\end{table}
{As in the commutative case, for the noncommutative SUSY case, we have a singularity in $r=0$. This singularity is independent of the value of $C$. We start with the solution with the positive sign, there are two possibilities, $C>0$ and $C<0$. The first case is a regular for $r>0$, for the second case we have a singularity in $r=C^2$. Taking the solution with the negative sign, again we have two cases. For positive $C$ the time component of the metric vanishes. Therefore, there is a singularity in $r = C^2$.   The case for negative $C$ is negative and therefore this case is discarded.

In order to exhibit the singularities, the Kretschmann invariant is calculated and it is given by}

\begin{eqnarray}
K&=&
\frac{1}{864 \sqrt[3]{6}  \left(\pm 1+\frac{C}{\sqrt{r}} \right)^{8/3}r^{22/3}}\\
&&\times\left[
\begin{array}{c}
3888\ 6^{2/3} C^4+ 216 C^3 \sqrt{r} \left(\pm 63\ 6^{2/3} -48 \sqrt{r} \left(C\pm\sqrt{r} \right)^{1/3}\right)\\
+27 C^2 \left(659\ 6^{2/3} r \mp1152 r^{3/2} \left(C\pm\sqrt{r} \right)^{1/3}+128 \sqrt[3]{6} r^{2} \left(C\pm\sqrt{r} \right)^{2/3}\right)\\
\pm24 C r^{3/2}\left( 431\ 6^{2/3} \mp1296 r^{1/2} \left(C\pm\sqrt{r} \right)^{1/3}+288 \sqrt[3]{6} r \left(C\pm\sqrt{r} \right)^{2/3}\right)\\
+32 r^2\left(108 \sqrt[3]{6} r \left(C\pm\sqrt{r} \right)^{2/3}+71\ 6^{2/3} \mp324 r^{1/2} \left(C\pm\sqrt{r} \right)^{1/3}\right)
\end{array}
\right]\nonumber\\
&+&\frac{\vartheta}{960 \left(\pm 1+\frac{C}{\sqrt{r}}\right)^{3}r^{15/2}}\nonumber\\
&&\times\left[
\begin{array}{c}
69984 C^5+ 972 C^4 \sqrt{r}\left(\pm281-16 \sqrt[3]{6} r^{1/2} \left(C\pm\sqrt{r} \right)^{1/3}\right)\\
+27 C^3 r\left(15355\mp 1984 \sqrt[3]{6} r^{1/2} \left(C\pm\sqrt{r} \right)^{1/3}\right)\pm13952 r^{5/2}\\
+9 C^2 r^{3/2}\left(\pm33767-7488 \sqrt[3]{6} r^{1/2} \left(C\pm\sqrt{r} \right)^{1/3}\right)\\
+84 C r^2\left(1267 \mp 432 \sqrt[3]{6} r^{1/2} \left(C\pm\sqrt{r} \right)^{1/3}\right)-6912 \sqrt[3]{6} r^{3} \left(C\pm\sqrt{r} \right)^{1/3}
\end{array}
\right].\nonumber
\end{eqnarray}
{From this equation, we can see that there are singularities for $r=0$ and $r=C^2$. 
Moreover, from the Kretschmann  invariant we conclude (as in the commutative case) that the singularity  in the SUSY region remains.}

\section{Concluding remarks}\label{final}
In this paper we have combined the ideas of noncommutative black holes \cite{julio1} and  SUSY black holes \cite{LopezDominguez:2007zz} to give a proposal for a noncommutative SUSY black hole. The approach is based on using the diffeomorphism between the KS and Schwarzschild metrics and introducing noncommutativity and supersymmetry to the KS-WDW equation. Furthermore, by solving the semiclassical approximation, we can reconstruct the metric for the noncummutative SUSY metric for the Schwarzschild black hole. Moreover, we calculate the Kretschmann scalar from which we conclude that for $\vartheta=0$ we recover the result for the SUSY case\cite{julio1}. This gives  us confidence that this is a noncommutative generalization of ``SUSY Schwarzschild metric" \cite{julio1}. We also see that the singularities for $r=0$ and $r=C^2$ are not removed by combining supersymmetry with noncommutativity \footnote{Even if in the metric it seems that for $r=C^2$ and taking the positive sign, the time component of the metric is non zero.}. Therefore, the behavior of this noncommutative SUSY black hole is qualitatively the same as for the proposed SUSY Schwarzschild black hole\cite{julio1}. 

{Alternatively, we could have started with the NC-WDW and constructed the SUSY generalization. In particular, supersymmetric versions of noncommutative quantum mechanical models have been constructed \cite{gamboa} with the noncommutative deformation  introduced using the Bopp shift. One characteristic feature of these models, is that angular momentum type interaction are introduced. Therefore, it can be expected that in the SUSY deformation of the NC-WDW equation, new angular momentum type interaction might appear.}

{Finally, the procedure used in this paper} can be applied to other black holes, in particular the (anti)de-Sitter black hole \cite{julio2}. These ideas are under research and will be reported elsewhere.

\section*{Acknowledgements}
This work is supported by CONACYT grant 258982. M. S. is supported by {CIIC-071/2022}, Mena-Barboza is partially  supported by Adler.

\end{document}